\documentstyle[preprint,prl,oldlfont,aps]{revtex}
\begin{document}
\draft
\title{Cooperativity of Protein Folding and the Random-Field Ising Model} 
\author{A. M. Gutin, V. I. Abkevich, and E. I. Shakhnovich}
\address{Harvard University, Department of Chemistry\\
12 Oxford Street, Cambridge MA 02138}
%\date{\today}
\maketitle
\begin{abstract}
The relation between cooperativity of protein folding and the Random-Field Ising Model (RFIM) is established. Generalization of the Imry-Ma argument predicts cooperative folding transition for small heterogeneity of the interactions stabilizing the native structure. Monte Carlo simulation of a lattice model shows that starting from some finite heterogeneity folding transition is not cooperative and involves formation of domains.
\end{abstract}
\pacs{PACS numbers: 87.10.+e, 82.20.Db, 75.10.Hk, 05.50.+q, 64.60.Cn}

%%%%%%%%%%%%%%%%%%%%%%%%%%%%%%%%%%%%%%%%%%%%%%%%%%%%%%%%%%%%%
%%%%%%%%%%%%%%%%%%%%%%%%%%%%%%%%%%%%%%%%%%%%%%%%%%%%%%%%%%%%%
\narrowtext

The statistical mechanics of protein folding begins from the pioneering work of Go and coworkers \cite{Go}. In this study a simple lattice model able to fold into a unique 3D structure was introduced. It has been demonstrated \cite{Go,BW87,SG89} that the folding transition in this model is of the first order or, in the terminology of protein science, cooperative. 

The Hamiltonian of the model can be written as follows
\begin{equation}
H=-b \sum_{i<j} \Delta (r_i^0,r_j^0) \Delta (r_i, r_j)
\end{equation}
where indexes $i$ and $j$ numerate monomers along a protein chain, $r_i$ is spatial lattice coordinates of monomer $i$ in a current conformation, and $r_i^0$ is spatial lattice coordinates of monomer $i$ in some target quenched conformation. Only self-avoiding conformations are allowed. $\Delta  (r_i, r_j)=1$, if monomers $i$ and $j$ are in contact, that is they are neighbors on a lattice, and $\Delta  (r_i, r_j)=0$, otherwise. 

The meaning of the Hamiltonian is simple; the monomers $i$ and $j$ attract each other with the energy constant $b>0$ if they are in contact in the target conformation, otherwise they do not interact. It is clear that the ground state of the chain corresponds to $r_i=r_i^0$, that is the target conformation has the lowest energy. Thus, the target conformation should be associated with the native state of a protein. Correspondingly, the contacts present in the target conformations will be referred to as to native contacts. Simulation of the model by means of Monte Carlo on a lattice showed that starting from random unfolded conformations a chain is able to fold spontaneously into the target conformations \cite{Go}. Moreover, upon change in the parameter $b$ a first order transition is observed between unfolded state corresponding to a number of coil conformations with a few contacts and the folded state corresponding to fluctuations around the target conformations. Thus, this simple model demonstrates a behavior characteristic for small globular proteins.

There are two main approximations in the model. First, all the interactions between the monomers not forming a contact in the native conformation are neglected. This approximation is based on the assumption that protein amino acid sequences are evolutionary selected to provide a low energy for the native conformations. It has been shown that low energy of the native conformation is a necessary and sufficient condition for a rapid folding to a stable structure \cite{BW87,SG93,SSK,GAS95}. It is expected that the sequences selected by the requirement to have low energy in the native conformation exhibit a first order phase transition between the unfolded state with only a few native contacts in it and the folded state close to the native conformations. The non-native contacts contribute substantially to the unfolded state only determining the compactness of the unfolded state. If non-native contacts are attractive enough on average, then the unfolded state is a compact globule, otherwise it is a coil state. 

Second approximation in the model is that all the native contacts are assumed to be of the same strength. In the present letter we relax this approximation and study the effect of heterogeneity of the native contacts on the folding transition. It appears that the folding transition for small heterogeneity of the native interactions is of first order, but starting from some finite heterogeneity folding transition is not cooperative and involves formation of domains.

The Hamiltonian we study is the following
\begin{equation}
H=- \sum_{i<j} b_{ij} \Delta (r_i^0,r_j^0) \Delta (r_i, r_j)
\end{equation}
where $b_{ij}$ are independent random quenched parameters with the same Gaussian distribution with the mean $b$ and the dispersion $\delta$. If $\delta=0$, then we have the original model described by the Hamiltonian (1). It was mentioned above that in this case there is first order folding transition. What happens if $\delta \ne 0$?

To approach this problem one can consider the following analogy between the studied system and the RFIM \cite{FA}. Each native contact between monomers $i$ and $j$ can be mapped onto a spin $s_k$ in the RFIM so that $s_k=2 \Delta (r_i, r_j)-1$. As a result of this mapping, the folded state corresponds to ferromagnetic oder in the up direction, and the unfolded state corresponds to ferromagnetic oder in the down direction. Accordingly, the attraction parameters $b_{ij}$ corresponds to a local external magnetic field $h_k$. Finally, the counterpart for the exchange interactions between spins responsible for the ferromagnetic order is established by the fact that formation of a native contact makes easier formation of another neighboring native contacts. The effective interaction between neighboring contacts results in free energy cost of interface between the folded and unfolded phases. This free energy cost corresponds to the surface energy of a domain wall in the RFIM.

In addition to the effective interaction between the native contacts which are neighbors in the native structure there is another kind of interactions due to polymeric structure of the system. Indeed, formation of the native contact between monomers $i$ and $j$ with $i<j$ makes more favorable formation of the native contact between monomers $i_1$ and $j_1$ if $i_1>i$ and $j_1<j$. It should be mentioned that the polymeric effect results in effective interactions between contacts which may be distant from each other in the native structure.

The described analogy between the present model and the RFIM brings the idea to apply the domain argument of Imry and Ma \cite{IM} to analyze the effect of randomness of $b_{ij}$ on the thermodynamics of protein folding. First, we consider the stability of the folded state with respect to melting of some small part of the native structure. The average attraction $b$ between interacting monomers should be chosen so that the unfolded and folded states have equal stability, that is the system is at transition point. If $R$ is the size of the melted domain, then the surface energy cost is of order $R^{d-1}$ in $d$ dimensions, and the gain in the energy is of order $\delta R^{d/2}$. The latter gain is due to the fact that statistically the contacts from the melted domain are less stable than the native contacts on average. 

The surface energy cost and the statistical gain in the free energy are similar to those in the RFIM. In addition, there is entropic cost due to polymeric bonds. A domain of size $R$ consists of a number of subchains with a typical number of monomers in one subchain $g$ (Fig. 1a). $g$ can be estimated from the condition that the ideal size of a subchain is of oder of the domain size: $ag^{1/2} \sim R$, where $a$ is the size of a lattice bond. The number of such subchains in one domain is $n \sim R^d/g \sim R^{d-2}$. The loss of entropy upon melting of the domain is due to the fact that the both ends of each subchains are still fixed. The loss of entropy due to fixing of one end is of order $\ln ((R/a)^d)$, so the overall loss of entropy per domain due to polymeric bonds is of order $R^{d-2} \ln R$. 

Comparing this loss of entropy at large $R$ with the surface energy cost one can notice that at any dimension polymeric nature of the system is irrelevant. This shows that the described analogy with the RFIM is exact when the stability of the folded state with respect to small disorder in the energies of the native contacts is considered. Thus, one can immediately conclude that the lower critical dimension for protein folding is $d_c=2$, so that in three dimensions the folded state is stable with respect to small heterogeneity in the energies of the native contacts \cite{IM}. Moreover, there must be some finite critical value of the heterogeneity $\delta_c$ above which the unfolding transition is not of the first order anymore and involves non-cooperative melting of domains. It should be noticed that in two dimensions even small disorder should destroy the first order folding transition, so that a precaution should be used in interpretation of lattice simulations on two-dimensional lattices.

Up to now the stability of the folded state with respect to the disorder was analyzed. The analysis of the stability of the unfolded state deserves a separate consideration. The first step is to estimate the free energy of formation of one folded domain of size $R$. The surface energy cost and the statistical energy gain are the same as in the considered case of melting of a domain on the background of the folded state. However, the loss of entropy due to polymeric bonds is very different. Now the ends of loops of the length about $N/n$, where $N$ is the length of the whole chain, are fixed (Fig. 1b). The corresponding loss of entropy is of order $n \ln (N/n) \sim R^{d-2} \ln (N/R^{d-2})$. It is seen that for large $N$ the loss of entropy is very large compared to the surface energy cost, so that the unfolded state seems to be very stable with respect to disorder. 

In fact, this is not the case. The unfolded state is really stable with respect to a formation of one domain only. The situation changes drastically when formation of many domains simultaneously is considered (Fig. 1c). In contrast to the RFIM where there is no interaction between such domains, in the present system polymeric bonds result in entropic interaction between domains. The accurate estimate of the loss of entropy per domain due to polymeric bonds in the case of many domains gives the same result as for the melting of a domain on the background of the folded state. As a result, the folded and unfolded states are stable under the same conditions.

It should be mentioned that all the obtained conclusions are based on some heuristic arguments of Imry-Ma type. The rigorous prove analogous to that of Bricmont and Kupiainen \cite{BK} for the RFIM is still needed. Meantime it makes sense to check the conclusions by means of simulation of a lattice model. A cubic lattice was taken together with the standard Monte Carlo method \cite{VS} to simulate motion of a chain. The target conformation used in the simulation is shown in Fig. 2. This maximally compact conformation of a chain of $N=48$ monomers has $C_{max}=57$ contacts. Correspondingly, $C_{max}$ random numbers $e_{ij}$ with the Gaussian distribution with the zero average and unit dispersion were generated. Then we set $b_{ij}=b+ \delta e_{ij}$. For each of the values of $\delta$ the transition value of $b_{tr}(\delta)$ was estimated by means of long Monte Carlo simulations. The transition point was defined to have $C_{max}/2$ native contacts on average. Then, another long ($10^8$ Monte Carlo steps) simulation at $b=b_{tr}(\delta)$ was performed to collect the data for the histogram corresponding to the distribution of the number of the native contacts $C$. The result is presented in Fig. 3.

It is clearly seen that at $\delta =0$ the distribution of the number of the native contacts is bimodal with two picks. One pick at $C \approx 7$ should be associated to the unfolded state and the other pick at $C=C_{max}$ should be associated to the folded state. Thus, the system exhibits the behavior typical for a first order phase transition. For small $\delta$ the picture does not change qualitatively with increase in $\delta$; still bimodal distribution is observed but the picks shift to each other. Finally, at $\delta _c \approx 1.7$ the behavior changes qualitatively; at $\delta > \delta_c$ the distribution of $C$ is monomodal with a maximum about $C_{max}/2$ which implies that the folding transition is continuous. Thus, the results of the simulation of the lattice model are in accordance with the predictions of the heuristic arguments of Imry-Ma type.

An obvious application of the obtained results is related to the protein design problem. In order to design a stable and rapidly folding protein one has to provide low energy of a target conformations relative to the other misfolded conformations. Two parameters, $Z$-score \cite{BLE} and the ratio $T_f/T_g$ \cite{BW87} were proposed as a quantitative measure of the design quality. Both of the parameters are based on the approximation of the protein conformational space by the Random Energy Model (REM) \cite{Derrida} and, in fact, within the REM approximation these parameters are monotonic functions of each other. More important, both of the parameters do not depend on the dispersion $\delta$ of the energies of the native interactions and, therefore, they are not sensitive to the effects studied in the present letter. 

Thus, a design procedure based on $Z$-score or on the ratio $T_f/T_g$ results in some value of $\delta=\delta_0$. If $\delta_0 < \delta_c$, then the design procedure generates proteins having cooperative folding transition. However, if $\delta_0 > \delta_c$, then the design procedure generates proteins with thermodynamic domains. It should be noticed that in this case in order to stabilize the folded state one has decrease the temperature so that the folding rate becomes very slow. Interestingly, the design of model lattice proteins of moderate size \cite{AGS95} based on $Z$-score actually produces chains with the thermodynamic domains which implies that $\delta_0 > \delta_c$ in this model. Thus, in order to design a protein with cooperative folding transition one has to control value of $\delta$. An example of the improved design algorithm which tends to minimize $\delta$ is studied elsewhere \cite{AGS96}.   

This work was supported by the David and Lucille Packard Funds.

\pagebreak

{\bf Figure Captions}

{\bf Fig.1} Stability with respect to formation of domains. Shaded areas correspond to the folded state. (a) Melting of a domain in the folded state. (b) Formation of a domain in the unfolded state. (c) Interaction between the formed domains in the unfolded state. 

{\bf Fig.2} Target conformation of a chain of 48 monomers on a cubic lattice. The conformation was obtain by folding of a homopolymer with attraction between its monomers.

{\bf Fig.3} Distribution of the number of the native contacts $C$ at different heterogeneity $\delta$.


\begin{thebibliography}{10}

\bibitem{Go}
H. Taketomi, Y. Ueda, and N. Go, Int. J. Peptide Protein Res. {\bf 7}, 445 (1975).

\bibitem{BW87}
J. D. Bryngelson and P. G. Wolynes, Proc.~Natl.~Acad.~Sci.~USA {\bf 84}, 7524 (1987).

\bibitem{SG89}
E. I. Shakhnovich and A. M. Gutin, Studia Biophysica {\bf 132}, 47 (1989).

\bibitem{SG93}
E. I. Shakhnovich and A. M. Gutin, Proc. Natl. Acad. Sci. USA {\bf 90}, 7195 (1993).

\bibitem{SSK}
A. Sali, E. Shakhnovich, and M. Karplus, J. Mol. Biol. {\bf 235}, 1614 (1994).

\bibitem{GAS95}
A. M. Gutin, V. I. Abkevich, and E. I. Shakhnovich, Proc. Natl. Acad. Sci. USA {\bf 92}, 1282 (1995).

\bibitem{FA}
S. Fishman and A. Aharony, J. Phys. C {\bf 12}, L729 (1976).

\bibitem{IM}
Y. Imry and S. Ma, Phys. Rev. Lett. {\bf 35}, 1399 (1975).

\bibitem{BK}
J. Bricmont and A. Kupiainen, Phys. Rev. Lett. {\bf 59}, 1829 (1987).

\bibitem{VS}
P. H. Verdier and W. H. Stockmayer, J. Chem. Phys. {\bf 36}, 227 (1962).

\bibitem{BLE}
J. U. Bowie, R. Luthy, and D. Eisenberg, Science {\bf 253}, 164 (1991).

\bibitem{Derrida}
B. Derrida, Phys. Rev. Lett. {\bf 45}, 79 (1980).

\bibitem{AGS95}
V. I. Abkevich, A. M. Gutin, and E. I. Shakhnovich, Protein Sci. {\bf 4}, 1167 (1995).

\bibitem{AGS96}
V. I. Abkevich, A. M. Gutin, and E. I. Shakhnovich, Folding \& Design, in press.


\end{thebibliography}
\end{document}